# Programmable Quantum Anomalous Hall Insulator in Twisted Crystalline Flatbands

Wenxuan Wang[1]*, Yijie Wang[1]*, Zaizhe Zhang[1], Zihao Huo[1], Gengdong Zhou[1], Shu Zhang[2], Kenji Watanabe[3], Takashi Taniguchi[4], Xiaoxia Yang[2], Qing Dai[2,5], X.C. Xie[1,6,7], Kaihui Liu[8]†, Zhida Song[1,9] † and Xiaobo Lu[1,9] †

[1]International Center for Quantum Materials, School of Physics, Peking University, Beijing 100871, China
[2]CAS Key Laboratory of Nanophotonic Materials and Devices, CAS Key Laboratory of Standardization and Measurement for Nanotechnology, CAS Center for Excellence in Nanoscience, National Center for Nanoscience and Technology, Beijing 100871, China
[3]Research Center for Electronic and Optical Materials, National Institute for Material Sciences, 1-1 Namiki, Tsukuba 305-0044, Japan
[4]Research Center for Materials Nanoarchitectonics, National Institute for Material Sciences, 1-1 Namiki, Tsukuba 305-0044, Japan
[5]School of Materials Science and Engineering, Shanghai Jiao Tong University, Shanghai 200240, China
[6]Interdisciplinary Center for Theoretical Physics and Information Sciences, Fudan University, Shanghai 200433, China
[7]Hefei National Laboratory, Hefei 230088, China
[8]State Key Laboratory for Mesoscopic Physics, Frontiers Science Centre for Nano-optoelectronics, School of Physics, Peking University, Beijing 100871, China
[9]Collaborative Innovation Center of Quantum Matter, Beijing 100871, China

*These authors contribute equally to this work.
†Corresponding author. E-mail: xiaobolu@pku.edu.cn; songzd@pku.edu.cn; khliu@pku.edu.cn

**The isospin flavors in condensed matters can be continuously broken, forming various symmetry-broken quantum states. In moiré crystals, the competition between different isospin configurations can be effectively tuned by the twist angles and staciking orders. Here we report twisted double rhombohedral-trilayer-gaphene as a new twisted crystalline flatbands system showing rich moiré dependent topological phenomena. In devices with small twist angles, programmable Chern insulators with Chern number $C = 3$ at integer moiré filling $\nu = 1$ have been observed. We have further revealed an exotic hidden order which can quench the Chern insulator as well as multiple first-order transitions between different symmetry-broken phases. Interestly, in the device with a slightly larger twist angle, multiple Chern insulators with $C = 1$ at fractional moiré fillings including $\nu = 1/4$, 1/3 and 1/2 have been observed, whereas the Chern insulator at $\nu = 1$ is absent. Our study demonstrated the twisted flatbands form rhombohedral-multilayer-graphene as a new platform to study tunable high Chern insulators as well as new devices for quantum storage and computation.**

Moiré crystals featured with the tunability of twist angle and stacking orders have led to a paradigm shift in quantum band engineering. Graphene based moiré systems have been an important material family for the study of strongly correlated physics since the discovery of magic angle graphene[1-3]. So far, there are two major technical routes towards realizing strongly correlated phenomena in moiré graphene. One family is composed of different twisted graphene systems with monolayer or Bernal-stacked few-layer graphene as building blocks. Represented by magic-angle bilayer graphene, such family also includes magic angle twisted multilayer graphene[4-9], twisted mono-bilayer graphene[10-13], twisted double bilayer graphene[14-17], and other combinations[18-20]. In these systems, the formation of the flat bands relies on the cooperative effect of the moiré potential and interlayer hopping, which can be well described by the continuum model[1, 21-24]. The other family mainly contains the rhombohedral graphene/hBN moiré systems. Unlike twisted graphene, the flat bands in rhombohedral graphene originate from its intrinsic low-energy excitations[25-34]. When aligned with hBN, the resulting moiré superlattice further folds the original energy bands, leading to similar flat moiré bands with twisted graphene[35-44].

Compared with above two systems, twisted rhombohedral graphene is simultaneously modulated by the periodic moiré potential, interlayer hopping and the intrinsic flat bands in the low-energy region, making it a promising new moiré graphene system with significantly enhanced strong correlations. Notably, recent observations of a variety of symmetry-breaking quantum states in the rhombohedral graphene/hBN superlattice, such as fractional Chern insulators and superconductivity[36,38-43], have drawn considerable attentions. However, studies on correlated quantum phenomena in twisted rhombohedral graphene remain scarce with a few experimental attempts been reported[45-46]. The main challenge is that rhombohedral stacking is a metastable phase, making it susceptible to structural phase transitions during multiple stacking processes.

In this work, we have demonstrated twisted rhombohedral trilayer graphene (TRTG) a realizable flat-band system. Rhombohedral stacked graphene is a thermodynamically metastable structure, which tends to relax into the Bernal-stacked configuration during the stacking process. Hence, we have developed a high-throughput fabrication method to overcome the challenge ((Extended Data Fig. 1). As shown in Fig.1a, we have successfully fabricated a dual gated TRTG device which contain three independent regions with twist angle $\theta$ = 1.20° (device D1), $\theta$ = 1.19° (device D2) and $\theta$ = 1.17° (device D3) and another device which contains a region with twist angle $\theta$ = 1.61° (device D4). The Raman spectroscopic results from these three regions clearly show their rhombohedral stacking characteristics (Fig. S1d). Our subsequent results are mainly focused on device D1 and device D4.

We have employed continuum model to obtain the non-interacting band structure at $\theta$ =1.2° and $\theta$ =1.6°, where displacement field modulated central flat bands can be observed(Fig.1b-d). At $\theta$ =1.2°, when the displacement field is zero, the intwined central flat conduction band and valence band are gapped from the adjacent dispersive bands (Fig.1b). With displacement field applied, an energy gap opens at the charge neutral point while simultaneously shifting the central flat conduction band to higher energies and the valence band to lower energies. This energy shift progressively reduces the separation between the central flat bands and their neighboring dispersive bands, ultimately leading to band touching at critical field strengths (Fig.1c). Notably, Well-defined inter-band gaps separate the second and third dispersive bands in both the conduction (electron) and valence (hole) regions. The Chern number of the flat conduction band remain $C$ = 3 with or without displacement field. At $\theta$ =1.6°, the conduction band of the central flat band manifests as a flat band in the low-energy region and transitions into a highly dispersive band in the high-energy region. Notably, as the displacement field increases, the band gap of the central flat bands initially closes and subsequently reopens, accompanied by a transition in Chern number from $C = 0$ to $C = 3$ (Fig. S3). Moreover, the central flat bands maintain isolation from adjacent dispersive bands under applied displacement fields.

Fig.1g-h displays the longitudinal resistivity $\rho_{xx}$ as a function of moiré filling factor $v$ and displacement field $D/\varepsilon_0$ for device D1 and D4, respectively, measured at $T = 1.5$K. The moiré filling factor $v$ is calibrated by magneto transport measurement (Supplementary Information). In device D1, Single particle features at $v = 0$, $\pm4$ and $\pm12$ align precisely with non-interacting band calculations (Fig. 1b, c). At charge neutrality ($v = 0$), the system transitions from metallic insulating behavior with increasing $D/\varepsilon_0$, indicating a gap opening at charge neutrality. At $v = \pm4$, the insulating states originate from the single-particle gap between the spin-valley degenerated central flat band and the adjacent dispersive band. At $v = \pm12$, the insulating states reflect the single-particle gap between the second and third dispersive bands. Beyond these single particle effects, correlated states also emerge at $v = 1$ and 2 at finite $D/\varepsilon_0$. Device D2 and D3 (similar twist angle to D1) exhibit nearly identical behavior (Fig. S4). For device D4, single-particle features at $v = 0$ and $\pm4$ similarly align with band calculations (Fig. 1d,e). At $v = 0$, the resistive state is initially suppressed and subsequently reappears with increasing $D/\varepsilon_0$, consistent with the closure and reopening of the central band gap. Resistive features between $v = 0$ and $v = 2$ further suggest interaction-driven phenomena. Crucially, all longitudinal resistance features of TRTG devices exhibit mirror symmetry about $D/\varepsilon_0$, confirming preservation of $C_{2x}$ symmetry without strain-induced or structural relaxation effects.

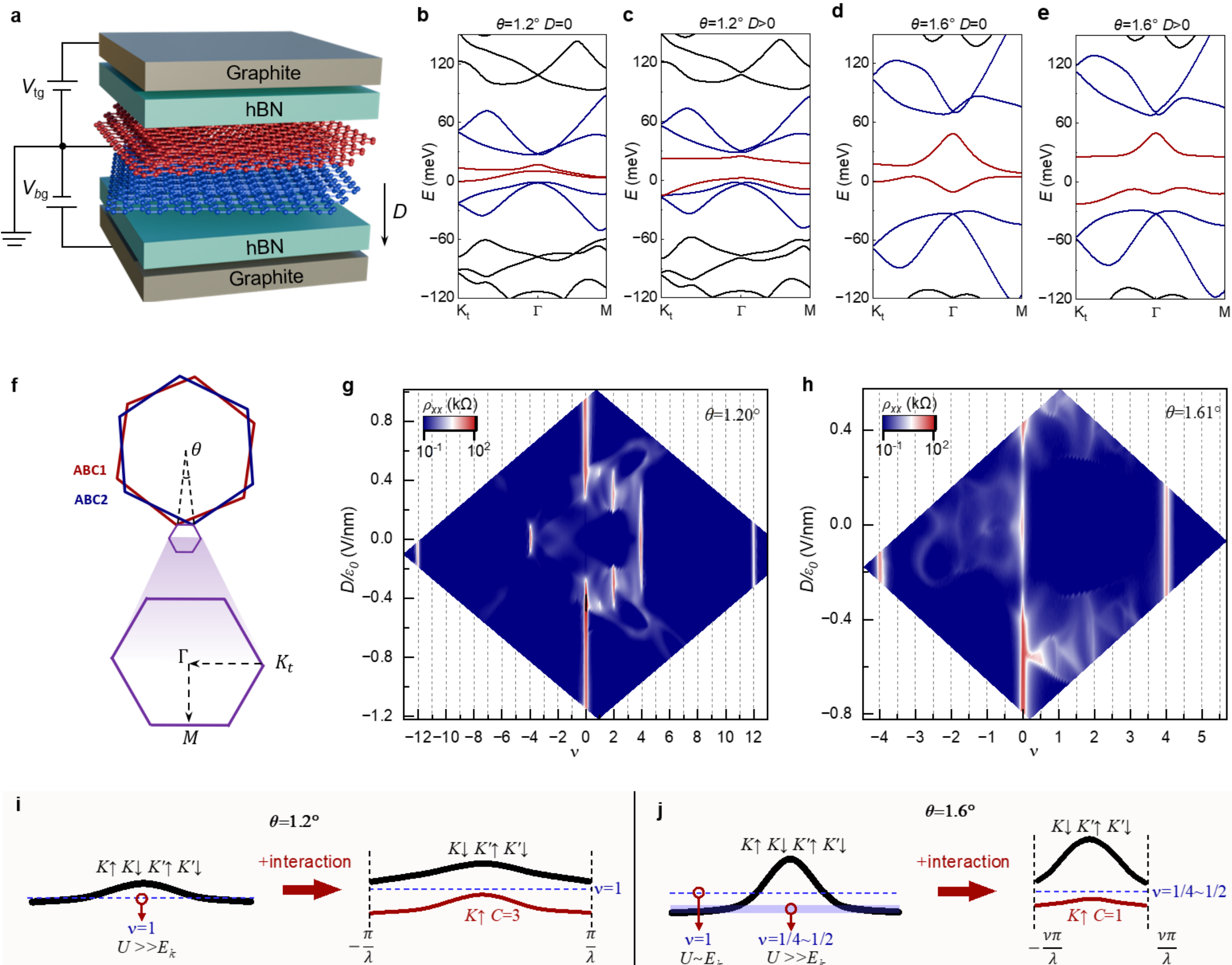

**Fig. 1|Topological flat band in the twisted rhombohedral trilayer graphene. a,** Schematic of the TRTG device, in which two rotationally misaligned rhombohedral stacked tri-layer graphene with twist angle $\theta$ are sandwiched between two graphite gates with hBN insulating spacer layers. **b-e,** Non-interacting band structure of TRTG at $\theta$=1.2° and $\theta$=1.6° with displacement field **(c, e)** and without **(b, d)** displacement field obtained from continuum model, where $\Delta$ = 6meV **(c)** or 9meV **(e)** is the potential difference between top and bottom graphene. **f,** Schematic of the Brillouin zone of TRTG, the dashed arrows denote the projection path. **g-h,** $\rho_{xx}$ as a function of filling factor $\nu$ and $D/\varepsilon_0$ measured at temperature $T$=1.5K for device D1 (**g**) and device D4 (**h**). **i-j,** Sketch of the flavor degenerated flat conduction band and the formation of Chern band driven by electron interaction for $\theta$=1.2° **(i)** and $\theta$=1.6° **(j)**.

The central flat conduction bands at $\theta$ =1.2° and $\theta$ =1.6° exhibit distinct topological characteristics. As shown in Fig. 1i, at θ = 1.2° and $\nu$ = 1, the band structure resides in the strong-correlation regime ($U >> E_k$). Upon introducing electron interactions, a correlated Chern gap opens, yielding a spin and valley polarized topological Chern insulator with $C$ = 3. At $\theta$ = 1.6°, the central conduction band is flat at bottom but becomes dispersive elsewhere. Consequently, the strong-correlation condition $U >> E_k$ is satisfied at fillings $\nu \approx$ 1/4 to 1/2, but not at $\nu$ = 1. Instability from electron interactions therefore spontaneously breaks translation symmetry and induce a correlated gap at these fractional fillings, leading to a redistribution of Berry curvature and a Chern number of $C = 1$.

Fig. 2a (b) shows $\rho_{xx}$ ($\rho_{xy}$) of device D1 as a function of $\nu$ and $D/\varepsilon_0$ measured at $B = 0.1$T and a base temperature of $T = 10$mK. Fig. 2e shows a zoom-in phase diagram around $\nu = 1$. At optimal $D$ field, the $\nu = 1$ state exhibits a minimum in $\rho_{xx}$ and a finite value in $\rho_{xy}$, indicating the emergence of a Chern insulator state. Fig. 2c displays the Landau fan diagram around $\nu = 1$. Notably, the $C = 3$ state which can survive to exact zero magnetic field also shows standard quantum anomalous Hall effect (Fig. 3d). The observed quantized Hall resistance and hysteretic behavior can emerge in a extensive range of displacement field and carrier density near $\nu = 1$ (Fig. S5). A further temperature dependent measurement shows a decent quantization can survive at a temperature up to $T \approx 1.7$ K. (Fig. S6a). Moreover, the preserved $C_{2x}$ symmetry also ensures symmetric manifestation of the quantum anomalous Hall state in both positive and negative $D$ fields (Fig. S5). Moreover, temperature-dependent magnetoresistance oscillations allow us to determine the effective mass $m^*$ (Fig. S6b-c). At $\nu = 1.54$ and $D/\varepsilon_0 = 0.436$V/nm, we extract an effective mass of $m^* = 0.55\pm0.02m_e$, where $m_e$ is the bare electron mass. Such value is approximately twice of that in twisted double bilayer graphene[15].

TRTG with $\theta$=1.20° also exhibits rich flavor polarized phenomena. In Fig. 2a, a phase boundary connecting the insulators at $\nu = 0$ and $\nu = 2$ at $D/\varepsilon_0 \approx \pm0.25$V/nm can be well distinguished. At lower $D$ field, the phase shows lower value of $\rho_{xx}$ which is in consistent with a normal metal phase. Such normal metal (NM) phase originates from the partially filled intertwined moiré bands which is evidenced by our calculations (Fig. 1b) and the signatures of Landau level crossings (Fig. S7). In the other side of the phase boundary corresponding to larger $D$ field, The SPM phase has been manifested by the hysteretic behavior of $\rho_{xx}$ when sweeping in-plane magnetic field back and forth (Fig. S8b-d). Moreover, the vanishing $\rho_{xy}$ in Fig. 2b as well as the two-fold degenerated quantum oscillations at finite perpendicular magnetic field (Fig. 2f and g) further confirm the spontaneous spin polarization of the phase. The valley polarization is absent since the concomitant orbital effect can result in large Hall signal. As the displacement field increases further, the system spontaneously breaks valley degeneracy, leading to a Chern insulator state at $\nu = 1$ and a quarter metal state at $\nu > 1$. As shown in Fig. 2f, such a transition occurs at $D/\varepsilon_0 > 0.42$V/nm near $\nu = 1$. The quarter metal (QM) phase is evidenced by the non-zero $\rho_{xy}$ as well as one-fold degenerated quantum oscillations shown in Fig. 2f and g. Importantly, as $B_\perp$ increase from 0.1T to 1T, the QM phase expands significantly in the $\nu$–$D$ diagram, since a higher $B_\perp$ field more effectively breaks valley degeneracy. Fig. 2h displays both $\rho_{xx}$ and $\rho_{xy}$ as a function of $D/\varepsilon_0$ at $\nu = 1$, quantitively showing three distinct phases including NM, SPM and a Chern insulator.

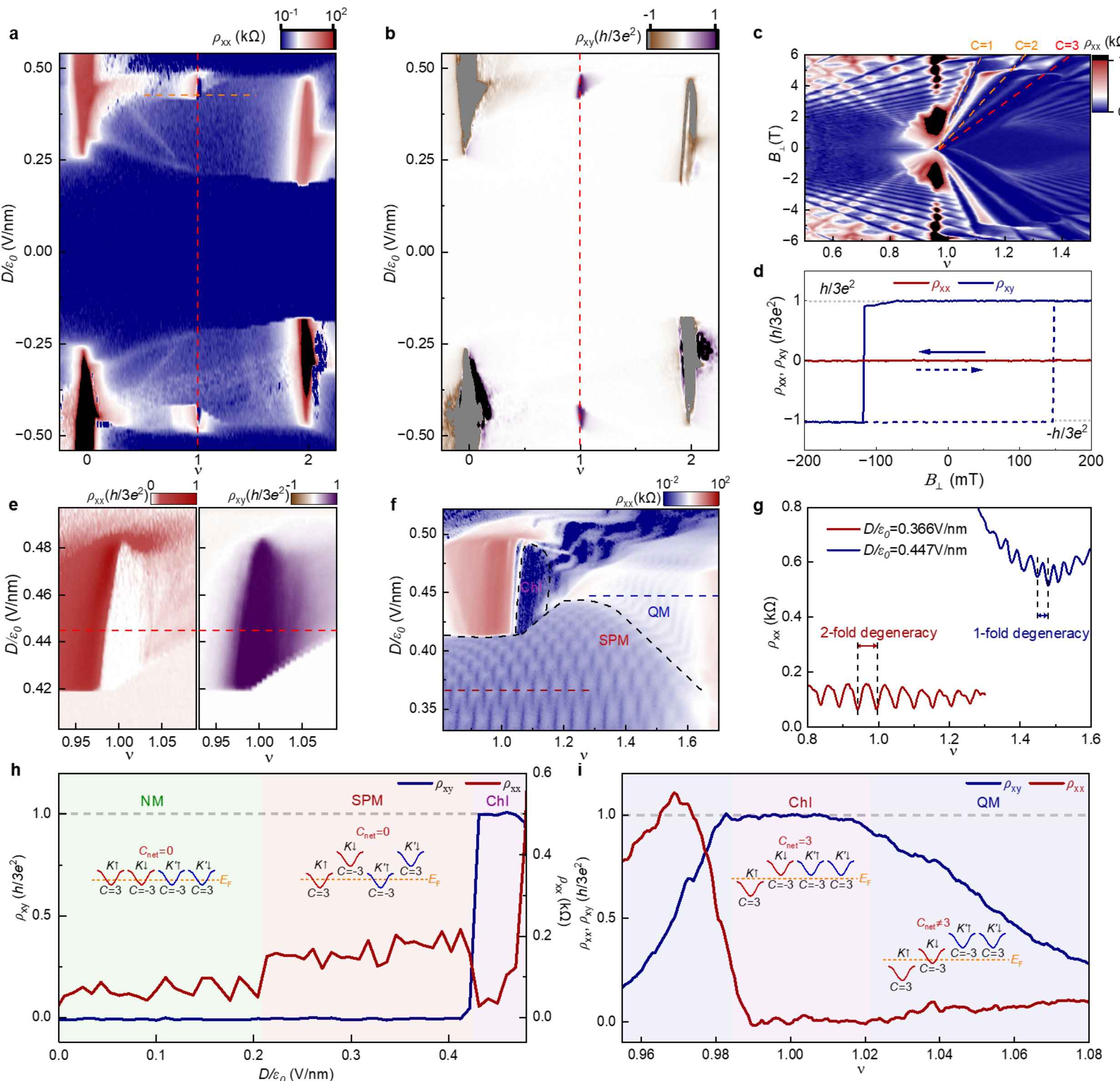

**Fig. 2|Chern insulators with $C$ = 3 and symmetry breaking correlated states in device D1 with $\theta$=1.20°. a-b,** Longitudinal resistivity $\rho_{xx}$ **(a)** and Hall resistivity $\rho_{xy}$ **(b)** as functions of $\nu$ and $D/\varepsilon_0$ measured at perpendicular magnetic field $B_\perp$ = 0.1T and temperature $T$ = 10mK. The red and orange dashed lines are the linecuts for **c** and **h** respectively. **c**, $\rho_{xx}$ as a function of $\nu$ and $B_\perp$ at $D/\varepsilon_0$ = 0.429V/nm. The dashed lines indicate the $\rho_{xx}$ minimal and strictly follow the Streda formula $C = h/e(\partial n/\partial B)$, where $C$ = 3. The red dashed line indicates the Chern insulator state, where the $\rho_{xx}$ minimal can persist to zero magnetic field. **d**, Magnetic hysteresis loop of $\rho_{xx}$ and $\rho_{xy}$ at $\nu$ = 1 and $D/\varepsilon_0$ = 0.429V/nm. Solid (dashed) lines correspond to sweeping $B_\perp$ from negative to positive (positive to negative). The arrows represent the $B_\perp$ sweeping direction. The black dashed lines indicate quantized value $h/3e^2$. **e,** $\rho_{xx}$ and $\rho_{xy}$ as functions of $\nu$ and $D/\varepsilon_0$ at perpendicular magnetic field $B_\perp$ = 0.1T near the Chern insulator state. The red dashed line is the linecut for **i**. **f,** $\rho_{xx}$ as a function of $\nu$ and $D/\varepsilon_0$ at $B_\perp$ = 1T. The black dashed lines indicate the phase boundaries. The blue and red dashed lines are the linecuts for **g**. QM, ChI, and SPM correspond to quarter metal, Chern insulator and spin polarized metal respectively. **g,** $\rho_{xx}$ as a function of $\nu$ at displacement field $D/\varepsilon_0$

= 0.366V/nm (red solid line) and $D/\varepsilon_0$ = 0.447V/nm (blue solid line), showing 2-fold degeneracy and 1-fold degeneracy respectively. **h,** $\rho_{xx}$ and $\rho_{xy}$ as functions of $D/\varepsilon_0$ at $v = 1$ and $B_\perp$ = 0.1T. As $D/\varepsilon_0$ increases, the spin degeneracy is firstly broken and the system turns from normal metal (NM) phase to spin polarized metal (SPM) phase, in which the net Chern number $C_{net} = 0$. As $D/\varepsilon_0$ further increases, the valley degeneracy is subsequently broken and the system turns into Chern insulator (ChI) phase with $C_{net} = 3$. **i,** Longitudinal resistivity $\rho_{xx}$ and Hall resistivity $\rho_{xy}$ as functions of $v$ at $D/\varepsilon_0$ = 0.446V/nm and $B_\perp$ = 0.1T. At $v$ slightly away from $v = 1$, the valley degeneracy is partially polarized and the quarter metal (QM) phase with $C_{net} \neq 3$ emerges.

Interestingly, first-order phase transitions which typically triggered by instabilities arising from interactions have been observed near both phase boundaries. We firstly focus on the phase transition near $v$=1 state. We define the Hall resistivity difference as $\Delta\rho_{xy} = \rho_{xy}(+) - \rho_{xy}(-)$, where "+" ("−") denotes positive (negative) sweep direction of $v$. Fig. 3a shows the $\Delta\rho_{xy}$ when sweeping $v$ from 0.92 to 1.04 back and forth at different $D$ field. The region rendered with red color in Fig. 3a indicates the hysteresis windows. Interestingly, when we extend the upper sweep range from $v_{max} = 1.04$ to $v_{max} = 1.10$, a significantly larger hysteretic window emerges (Fig. 3b). To further investigate the sweeping range dependent hysteretic behavior, we have measured the hysteretic loops with different values of $v_{max}$ at $D/\varepsilon_0$ = 0.421 V/nm. When $v_{max} < 1.058$, the hysteretic loops are almost independent of $v_{max}$ and the Chern insulator can be recovered with sweeping back $v$. However, when $v_{max} > 1.058$, the hysteretic window expands abruptly and the Chern insulator state cannot be recovered with sweeping back $v$. Fig. 3c and d show $\rho_{xy}$ as a function of magnetic field and $v$ measured at $D/\varepsilon_0$ = 0.421 V/nm. In sharp contrast with the Chern gap at $v = 1$ shown in Fig. 3c, the Chern gap in Fig. 3d corresponding to backwardly swept $v$ cannot persist to zero magnetic field. It seems that the system has a hidden switch $v_{switch}$. When $v_{max}$ increases beyond $v_{switch}$, the system will fall into a new phase, in which the valley degeneracy becomes more rigid. Although electrical switching of Chern insulator has been reported in other moiré systems[18, 42, 44], the hidden switch in TRTG has never been observed. The rigidity of the valley degeneracy with increased $v$ can also be evidenced by the detailed magneto transport near $v = 1$ (Fig. S9). When $v$ is increased, the Landau levels originate from $v = 0$ rather than $v = 1$, which indicates the closure of the correlated gap at $v = 1$ and the collapse of valley polarization.

To better understand the intriguing phase transitions, we have developed a Landau-Ginzburg model (Fig. 3f-h). For $v_{max} < v_{switch}$, the first-order transition of the valley polarization strength $\tau$ can be captured by a Landau-Ginzburg functional $f_1 = \alpha_2(v-v_1)\tau^2/2-\alpha_4\tau^4/4+\alpha_6\tau^6/6$ ($\alpha_{2,4,6}$>0), where $f_1$ is the free energy, $\alpha_{2,4,6}$ are constant coefficients (more details shown in method). The absolute values of $\alpha_{2,4,6}$ are not essential to our discussion, but their combination $\alpha_4^2/(4\alpha_2\alpha_6)$ determines another critical doping $v_2 = v_1 + \alpha_4^2/(4\alpha_2\alpha_6)$. When $v < v_1$, only the valley-polarized phase exists, when $v > v_2$, only the valley-symmetric phase exists (Fig. 3f-g). When $v$ is in between, the two phases can coexist, giving rise to the hysteresis.

For the transition with $v_{max} > v_{switch}$, the switch can be explained by assuming another hidden order $\zeta$, whose own hysteresis is given by Fig. 3h, with boundaries set by $v_0$ and $v_{switch}$. $\zeta$ is coupled to the valley order $\tau$ through a coupling $f_3 = \gamma\zeta^2\tau^2/2$ ($\gamma$>0) which describes a competition between the two orders. Once the sweeping window exceeds $v_{switch}$, the hidden order forms, which then sets an extra energy barrier to the valley symmetry breaking. During the back sweeping, supposing the competition $\gamma$ is strong, then the valley order can only form until the hidden order melts at $\nu_0$, so that the energy barrier is removed. This explains the larger hysteric window.

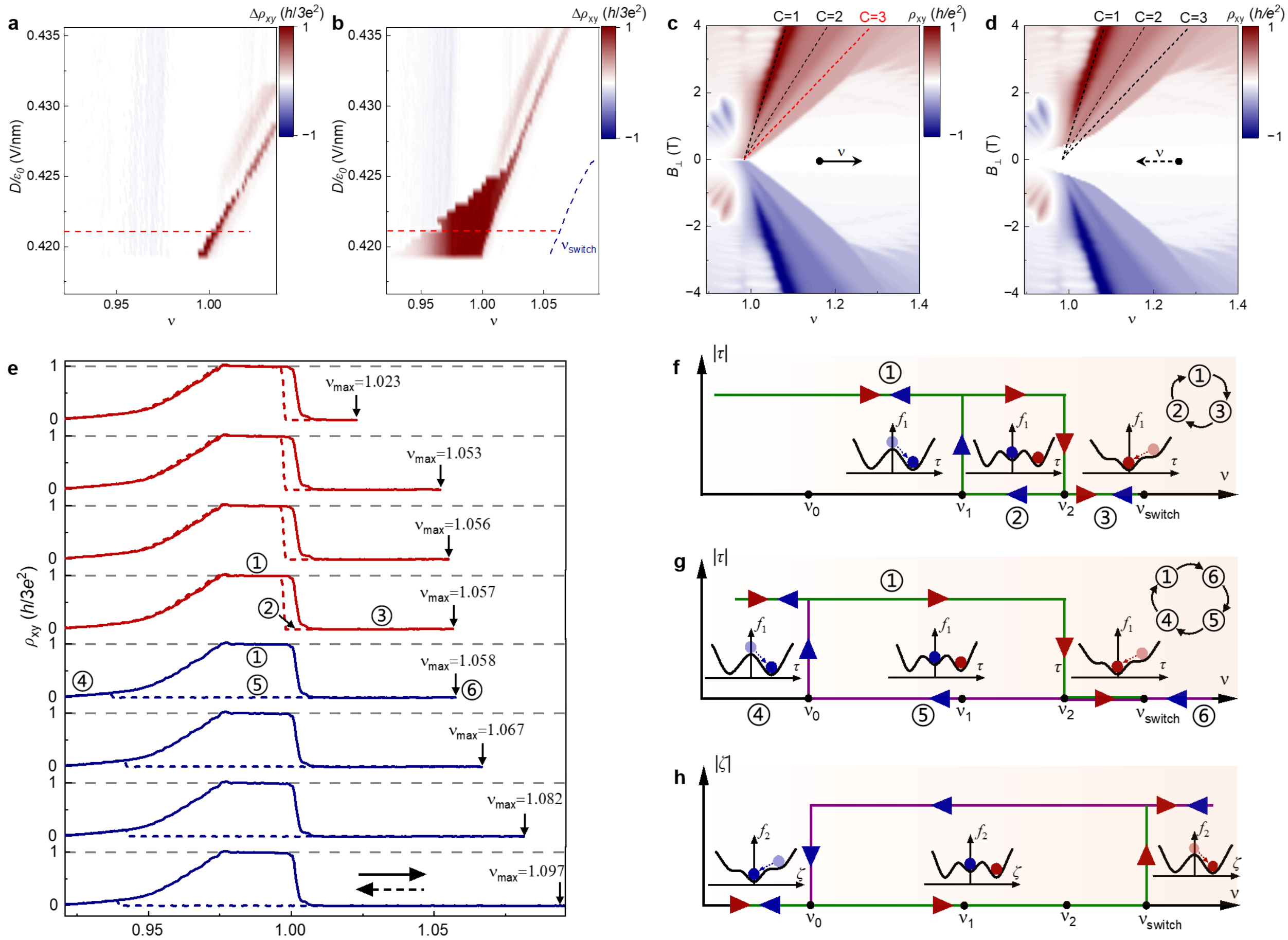

**Fig. 3| Programable Chern insulators in device D1with $\theta$=1.20°. a-b,** The difference of Hall resistivity $\Delta\rho_{xy}$ between positive and negative sweeping direction of $\nu$ as functions of $\nu$ and $D/\varepsilon_0$ with sweeping range $\nu$ = 0.92~1.04 **(a)** and $\nu$ = 0.92~1.10 **(b)**. The dark red region indicates the quenching of Chern insulator at negative sweeping direction. The blue dashed line indicates the hidden switch $\nu_{switch}$ as a function of displacement field $D$. **c-d**, Anti-symmetrized Hall resistance as a function of $\nu$ and $B_\perp$ with positive **(c)** and **(d)** negative sweeping direction at displacement field $D/\varepsilon_0$ = 0.421V/nm. The dashed lines indicate the $\rho_{xy}$ plateaus and strictly follow the Streda formula $C = h/e(\partial n/\partial B)$. The red dashed line indicates the Chern insulator state, where the $\rho_{xy}$ plateau can persist to zero magnetic field. The arrows represent the $\nu$ sweeping direction. **e,** $\rho_{xy}$ measured at $D$ = 0.421V/nm and $B_\perp$ = 0.1T with $\nu$ swept back (dashed line) and forth (solid line). The numbers above each arrow $\nu_{max}$ are the maximal of the $\nu$ sweeping range for each trace. The circled numbers represent the states shown in **f-g**. **f-g,** Sketches of the valley order $|\tau|$ dependency on $\nu$ near $\nu$ = 1 obtained from Landau-Ginzburg theory for small hysteresis **(f)** and large hysteresis **(g)**. The green (purple) solid lines denote that the hidden order $\zeta$ is (not) zero. The inlets exhibit the sketch of the dependence of the free energy $f_1$ on valley order $\tau$. **h,** Sketch of the hidden order $|\zeta|$ dependency on filling factor near $\nu$ = 1 obtained from Landau-Ginzburg theory. The inlets exhibit the sketch of the dependence of the free energy $f_2$ on hidden order.

Since the spin polarization can persist above $\nu_{switch}$, it is unlikely to serve as the hidden order. However, we note that with a large $D$ field, the flat band has a rather small gap to the remote bands, which can be verified both from the non-interacting band calculation, and from the experimental band gap closing at $\nu = 4$. Notably, the emergence of the hidden switch and the band gap closing at $\nu = 4$ take place at nearly identical displacement field $D/\varepsilon_0 \sim 0.4$V/nm (Fig. S7). This implies a stronger tendency to hybridizing the flat band with remote bands, which can trigger an orbital order that is associated with the spontaneous $C_{3z}$ symmetry breaking, and/or a band inversion that alters the band topology of the flat band. Through self-consistent Hartree-Fock calculations, such a valley-symmetric orbital order can indeed be found at $1<\nu<2$ (Fig. S2).

As observed first order transitions can also occur through sweeping $D$ field hysteresis. In Fig. S10a, the longitudinal resistivity difference is defined as $\Delta\rho_{xx}= \rho_{xx}(+) - \rho_{xx}(-)$. Here "+" ("−") represent sweeping $D$ field positively (negatively). Clear signal of $\Delta\rho_{xx}$ occurs near different phase boundaries as well when sweeping the $D$ field back and forth. Extended Data Fig. 10b shows the hysteretic behavior at $\nu = 2$ triggered by the phase transition between NM phase and SPM phase.The hysteretic behavior at $\nu = 1$ state (Extended Data Fig. 10c) manifests $D$ field induced phase transition from SPM to a Chern insulator with both spin and valley degeneracy broken. The non-volatile character of the Chern insulator, combined with its hysteretic transition on both $\nu$ and $D$ field, offers a promising platform for the low-power storage and computation transition based on the dissipationless edge state. A memory functionality is demonstrated in S11d by applying pulses of $D$ field and periodically setting and quenching the Chern insulator. Notably, a displacement pulse as small as 0.03V/nm can trigger the transition of Hall resistivity from zero to a quantum Hall plateau without magnetic field. Moreover, similar memory application can also be achieved by tunning $\nu$ pulses (Extended Data Fig. 10e-f).

Fig. 4a shows $\rho_{xx}$ of device D4 with $\theta = 1.6°$ as a function of $\nu$ and $D/\varepsilon_0$ measured at $B = 0.1$T and $T = 10$mK. Fig. 4b-c present zoom-in phase diagram of $\rho_{xx}$ and $\rho_{xy}$ near $\nu = 1/4\sim1/2$. At optimal $D$ field within this $\nu$ range, we observe simultaneous $\rho_{xx}$ minima and a finite $\rho_{xy}$ values, indicating the emergence of Chern insulator states. The linecut in Fig. 4d confirms Hall resistance quantization to a plateaus $\rho_{xy} = h/e^2$. Quantum anomalous Hall effects at $\nu = 1/4$, 1/3, and 1/2 provide further evidence for these topological states (Fig. 4e-g). As demonstrated by the Landau fan diagram in Extended Data Fig. 11, both $\rho_{xy}$ quantization and $\rho_{xx}$ minima persist down to $B_{\perp} = 0$ while rigorously obeying the Streda formula $C = h/e(\partial n/\partial B)$ for $C = 1$. Similar with previous reported twisted bi-trilayer graphene system[18], the Chern insulator appears at fractioanl filling while disappears at $\nu = 1$. Notably, the Chern insulator states here span a broad range of $\nu$, with maximum stability at commensurate fractional fillings $\nu = 1/4$, $\nu = 1/3$, and $\nu = 1/2$. Crucially, the nearly quantized anomalous Hall behavior can also emerge at incommensurate fillings (Extended Data Fig. 12).

To explore the Chern insulators stabilized at $\nu = 1/4\sim1/2$, we calculated the density of states (DOS) via continuum model. At optimal $D$ field ($\Delta = 9$ meV), the DOS shows pronounced concentration within $\nu = 1/4\sim1/2$, driving electronic instability and spontaneous symmetry breaking. The Chern number at $\theta = 1.6°$ and $\Delta = 9$meV obtained from the non-interacting band structure is $C = 3$ and contradicts to the experimentally observed $C = 1$. As shown in Fig. 4i, the Berry curvature is regularly distributed on the energy scale and does not concentrate on the bottom or top of the central conduction band. Thus, once a correlated gap at $\nu = 1/4\sim1/2$ opens, the integrated Berry curvature below the gap will be dramatically reduced. In rhombohedral graphene/hBN moiré superlattice, the Chern inslulator at both integer and fractional fillings has been observed at remote moire side [38-42]. However, their formation mechanism of such states in TRTG very likely differs due to distinct moiré potential responses under $D$ field. With applied $D$ field, the electrons are polarized to the top layers and nearly vanishes from the bottom layers (Fig. S13). Notably, there are still substantial electron density distribution at middle layers, where the moiré potential is very strong (Fig. S13c). Thus, we conclude that the formation of the Chern insulators at both commensurate and incommensurate fractional fillings is contributed from electrons away and near moiré potential, which is very different from the system of rhombohedral graphene/hBN moiré superlattice.

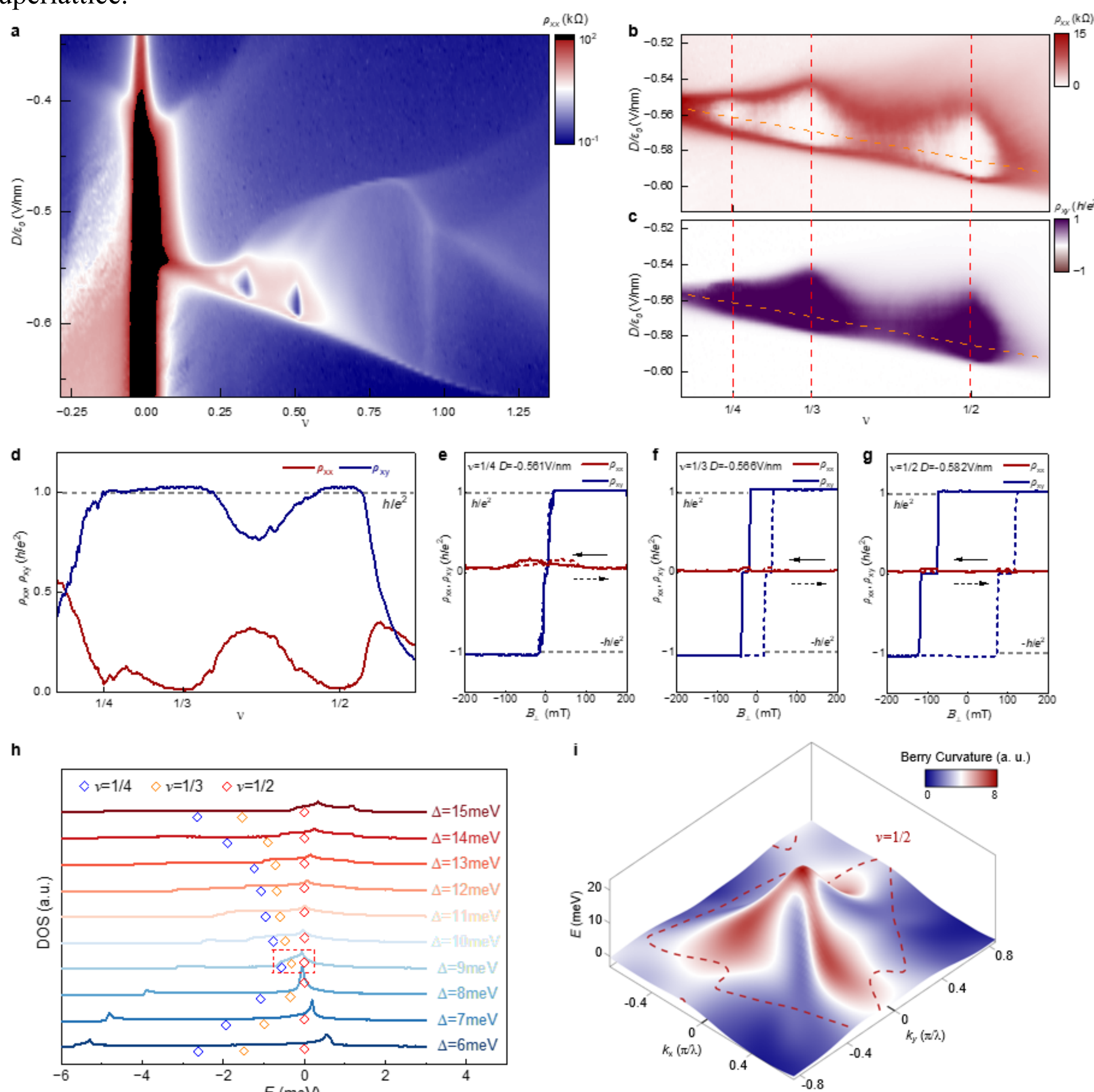

**Fig. 4| Chern insulators with $C$ = 1 at fractional moiré fillings in device D4 with $\theta$=1.61°. a,** $\rho_{xx}$ as functions of $\nu$ and $D/\varepsilon_0$ measured at perpendicular magnetic field $B_\perp$ = 0.1T and $T$ = 10mK. **b-c,** Symmetrized $\rho_{xx}$ **(b)** and anti-symmetrized $\rho_{xy}$ **(c)** as functions of $\nu$ and $D/\varepsilon_0$ at perpendicular magnetic field $B_\perp$ = ±0.1T. **d,** Symmetrized $\rho_{xx}$ and anti-symmetrized $\rho_{xy}$ as a function of $\nu$

measured along the trajectory indicated by the orange dashed line show in **b** and **c**. **e-g,** Hysteresis loop of symmtrized $\rho_{xx}$ and $\rho_{xy}$ at $\nu$=1/4, $D/\varepsilon_0$=0.561V/nm **(e)**, $\nu$=1/3, $D/\varepsilon_0$=0.566V/nm **(f)**, and $\nu$=1/2, $D/\varepsilon_0$=0.582V/nm **(g).** Solid (dashed) lines correspond to sweeping $B_\perp$ from negative to positive (positive to negative). The arrows represent the $B_\perp$ sweeping direction. The black dashed lines indicate quantized value $h/e^2$. **h,** The density of states for TRTG at $\theta$ =1.6° as functions of energy $E$ and $\Delta$ obtained from continuum model, where $\Delta$ is the potential difference between top and bottom graphene. The colored rhombus represent $E$ corresponding to $\nu$=1/4, $\nu$=1/3 and $\nu$=1/2. The red dashed rectangle represents that the density of states is concentrated on $\nu$=1/4~1/2 at $\Delta$=9meV. **i,** Berry curvature distribution of the central conduction band at $\theta$ =1.6° and $\Delta$=9meV obtained from continuum model. The red dashed line represents for the contour line of the Fermi energy where $\nu$=1/2. The Chern number obtained from integrating the Berry curvature of the whole central conduction band is $C$=3.

In conclusion, our work has demonstrated twisted rhombohedral graphene a promising platform to study topological correlated phenomena. As an example, TRTG exhibits rich interplay between different symmetry broken states and a programable Chern insulator. We stress that the twist angle in this system can fundamentally reconfigures the topological states: a small angle yields a $C$ = 3 Chern insulator at $\nu$ = 1, along with a hidden order and multiple first-order transitions, while a larger angle quenches the $\nu$ = 1 state, coinciding with the onset of $C$ = 1 Chern insulators at both commensurate fillings ($\nu$ = 1/4, 1/3 and 1/2) and incommensurate fillings. These Chern insulators, modulated by the twist angle, are very robust and remain quantized even at zero magnetic field. Given the high tunability of the twisted rhombohedral systems, further experiments of varying twist angles and layer numbers have great potential to boost the studies of strong correlation, novel band topology and possible superconductivity. During the preparation of the work, we noted a relevant study[47].

## Acknowledgements

This work was supported by the National Key R&D Program (Grant nos. 2022YFA1403500/02 and 2024YFA1409002), the National Natural Science Foundation of China (Grant Nos. 12274006, 12141401, 52025023 and 12427806), the JSPS KAKENHI (Grant nos. 21H05233 and 23H02052) and World Premier International Research Center Initiative (WPI), MEXT, Japan.

## Author Contributions

X.L., W.W. and K.L. conceived and designed the experiments; W.W fabricated the devices with help from Z.H.; W.W. performed the measurement with help from Z.Z.; W.W., K.L., X.X., Z.S., K.L. and X.L. analyzed the data; Y.W., GD.Z. and Z.S. performed the theoretical modeling; T.T. and K.W. contributed materials; W.W., Y.W., K.L., Z.S. and X.L. wrote the paper.

## Competing interests

The authors declare no competing interests.